\UseRawInputEncoding 
 \pdfoutput=1
\documentclass[journal=jacsat,manuscript=article]{achemso}

\usepackage[version=3]{mhchem} 

\author{Fan Peng}
\author{Cui Nie}
\author{Tingyu Xu}
\author{Junfang Sheng}
\email{jfsheng@ustc.edu.cn}

\author{Wei Chen}
\author{Wancheng Yu}

\author{Liangbin Li}
\email{lbli@ustc.edu.cn}

\affiliation[University of Science and Technology of China]
{National Synchrotron Radiation Laboratory, Anhui Provincial Engineering Laboratory of Advanced Functional Polymer Film, CAS Key Laboratory of Soft Matter Chemistry, University of Science and Technology of China, Hefei, 230026, China}
\title[An \textsf{achemso} demo]
  {Entanglement on nucleation barrier of \\ polymer crystal}

\abbreviations{IR,NMR,UV}

\begin{document}

\begin{tocentry}

Some journals require a graphical entry for the Table of Contents.
This should be laid out ``print ready'' so that the sizing of the
text is correct.

Inside the \texttt{tocentry} environment, the font used is Helvetica
8\,pt, as required by \emph{Journal of the American Chemical
Society}.

The surrounding frame is 9\,cm by 3.5\,cm, which is the maximum
permitted for  \emph{Journal of the American Chemical Society}
graphical table of content entries. The box will not resize if the
content is too big: instead it will overflow the edge of the box.

This box and the associated title will always be printed on a
separate page at the end of the document.

\end{tocentry}

\begin{abstract}
    We propose a theoretical approach to quantitatively account for the role of entanglement in the nucleation of polymer melts, which is the unique feature of polymer differentiated from small molecules. By performing molecular dynamics simulations, we obtain the nucleation barriers of polymer systems with different entanglement densities, which exhibits an opposite trend compared to the prediction of the classic nucleation theory (CNT). To amend the deficiency of the CNT in polymer crystallization, we introduce the entanglement free energy to reflect the role of entanglement in polymer nucleation. Specifically, the polymer nucleation not only involves free energies of monomers inside and on the surface of a nucleus as considered in the CNT, but also affects the entanglement network around the nucleus. Our theoretical approach provides a reasonable interpretation for the unsolved nucleation phenomena of polymers in simulations and experiments. 
\end{abstract}

\section{1 Introduction}

The connectivity of polymer chains is the unique feature that distinguishes polymers from small molecules \cite{ref1,ref2,ref3,ref4,ref5}, which is manifested as entanglement in bulk or concentrated solution. Entanglement is considered as a rheological concept well formulated by the tube model \cite{ref6,ref7,ref8,ref9}. However, how the entanglement affects phase transitions like polymer crystallization has not been well understood yet \cite{ref10,ref11,ref12,ref13,ref14,ref15,ref16}. Compared to small molecules, homogeneous nucleation of polymer crystal usually requires a large undercooling \cite{ref17}, implying that entanglement inhibits the nucleation. The nucleation of polymer systems with high entanglement density like polycarbonate (PC) is completely suppressed under conventional conditions \cite{ref18,ref19}. It is not clear whether this nucleation inhibition mainly stems from the entanglement-enhanced nucleation barrier or constrained chain dynamics. Therefore, the understanding of the role of entanglement in nucleation is critical for developing polymer crystallization theory, which is a longstanding challenge in polymer physics.

For decades, how the entanglement affects polymer crystallization has attracted great attention. Previous experiments revealed that the nucleation rate decreases with increasing the entanglement density \cite{ref16,ref20}. Yamazaki et al. proposed an empirical relation between the nucleation rate \( J \) and the entanglement density \( \upsilon_e \) as \( J \sim k\upsilon_e \) \cite{ref21}. Computer simulations demonstrated that the distribution of thickness of lamellar crystal statistically follows the local entanglement length \cite{ref22,ref23}. Yet, no quantitatively thermodynamic analysis of the entanglement on nucleation has been reported. Classic nucleation theory (CNT) regards the nucleation of polymers as the behavior of independent monomers, and subsequent modifications only consider the chain connectivity in amorphous structure on the fold surface of nucleus \cite{ref24}. Although the concept of reptation has been introduced into polymer crystallization model \cite{ref1,ref25}, we still lack a nucleation theory incorporating the unique connectivity of polymer chain explicitly. 

In this work, we have performed large-scale molecular dynamics (MD) simulations to study the role of entanglement in nucleation. Two methods are employed to obtain the nucleation barriers of systems with different entanglement densities, i.e., sampling nucleation events from the MD simulations directly, and theoretical calculation according to the CNT. Regrettably, the nucleation barriers obtained by these two methods show opposite trends with increasing the entanglement density. We ascribed this contradiction to the absence of the entanglement effect in the CNT. To amend this defect of the CNT, we introduce the entanglement free energy:

\begin{equation}
  G_\mathrm{z} = U_\mathrm{z} - TS_\mathrm{z}  \label{1}
\end{equation}

Where \( U_\mathrm{z} \) and \( S_\mathrm{z} \) are the entanglement energy and entropy, respectively. By incorporating the effect of \( G_\mathrm{z} \) into the CNT, we propose a new theoretical approach for the nucleation of polymers, which reaches a good agreement with the nucleation barriers obtained through sampling events in MD simulations.

\section{2 Methodology}

\subsection{2.1 Model and simulations}
All simulations are performed on open-source code LAMMPS \cite{ref26}. As we focus on the generic behavior of entangled polymer systems rather than microscopic properties, the widely used coarse-grained polyvinyl alcohol (CG-PVA) model with a fast crystallization rate is chosen to study polymer crystallization \cite{ref27}. Parameters of the CG-PVA model in reduced units originate from experimental data. Namely, the Boltzmann constant \( k_\mathrm{B} \), the mass of a monomer $m$, the length $\sigma$, and the time $\tau$ are reduced to 1. The bond length is \( b_\mathrm{0} = 0.5\sigma \). The reduced temperature $T$ = 1 corresponds to a real temperature 550 K, $\sigma$ = 1 corresponds to 0.52 nm, and $\tau$ = 1 is about 3.5 ps. The bond interaction is approximated by a harmonic potential,
\begin{equation}
  U_\mathrm{bond}(r) = \frac{1}{2} \,k_\mathrm{bond}(r-b_0)^2 \label{2}
\end{equation}

where \( k_\mathrm{bond} = 2704k_\mathrm{B}T/\sigma^2 \). The angular interaction is represented by a tabulated potential, and the non-bonded interaction is given by Lennard-Jones 9-6 potential, 
\begin{equation}
  U_\mathrm{nonb} = \varepsilon\left[\left(\frac{\sigma_0}{r}\,\right)^9 - \left(\frac{\sigma_0}{r}\,\right)^6\right] \label{3}
\end{equation}

where \( \sigma_\mathrm{0} = 0.89\sigma \), and \(\varepsilon = 1.511k_\mathrm{B}T\). The potential is truncated and shifted to 0 at the minimum of \( r_\mathrm{c} = 1.02\sigma \), which makes the non-bonded potential purely repulsive. The pressure $P$ = 8 is adopted to compensate for the attractive interaction, through which the crystallization is accelerated. 

By using the CG-PVA model, a series of polymer melts with different entanglement densities are generated. Firstly, simulation boxes with different initial side lengths \( L_\mathrm{0}\) (see Table S1 in SI) are created. Then, 200 single equilibrated chains with 1000 monomers ($N$ = 1000) are randomly placed into a simulation box. 
This step is referred to the \( place \) shown in Figure 1. In this way, a non-equilibrium polymer system with an initial entanglement density is prepared. The initial entanglement density relies on the magnitude of \( L_\mathrm{0}\). After the place, the simulation box is gradually compressed to a final side length $L$ = ca. 45 in a time duration of 2000$\tau$, during which the pressure of the polymer system is adjusted to $P$ = 8. However, during the compression, there would be some unreasonably local conformations in the polymer system. To eliminate these unreasonably local conformations, chains are relaxed for 500$\tau$ with both their ends being fixed, which assures that no obvious disentanglement occurs in the system during the relaxation. 
With the above procedures being completed, polymer systems with different entanglement densities are obtained. Note that for the largest simulation box with an initial side length \( L_\mathrm{0}\) = 600, as the space is large enough for the placement of chains, chains are regularly arrayed in the box, and keep wholly separated and thus unentangled. Naturally, the entanglement density in this case is the lowest we studied in this work. More details about the entire modeling process are given in S1 of the SI. 
The relationship between the density of the initial system before compression (\( \rho_\mathrm{place} \)) and the entanglement length in melt systems (\( \left \langle  N_{\mathrm{e0}}\right \rangle\)) is plotted in Figure 1. The Z1-code is used to identify the primitive path of chains, and measure the entanglement length of melts \( \left \langle  N_{\mathrm{e}}\right \rangle\) in the early stage of nucleation \cite{ref28,ref29,ref30,ref31}. The average entanglement length \( \left \langle  N_{\mathrm{e0}}\right \rangle\) of prepared polymer systems ($t$ = 0$\tau$) ranges from 267 to 17. \( \left \langle  N_{\mathrm{e0}}\right \rangle\) of the equilibrium melt is about 16.4 \cite{ref32}. 

\begin{figure}
  \centering
  \includegraphics[width=12cm,height=7.12cm]{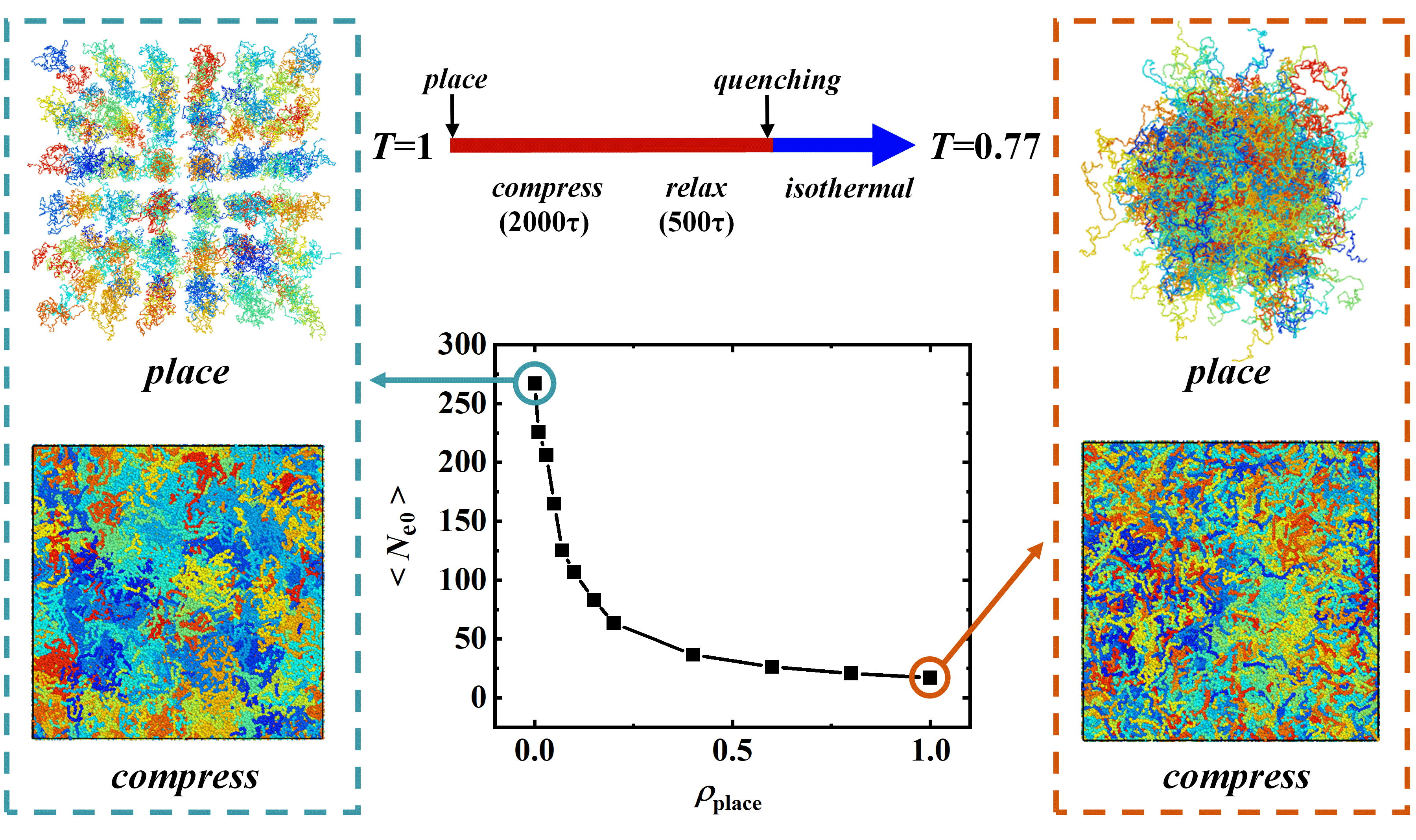}
  \caption{Schematic diagram of the modeling process, and snapshots of polymer systems with \( \left \langle  N_{\mathrm{e0}}\right \rangle\) = 267 and \( \left \langle  N_{\mathrm{e0}}\right \rangle\) = 17. Here, “\( place \)” means that randomly picking 200 points inside the simulation box first, and then laying the pre-prepared equilibrated chains with their center of mass on these picked points. Especially, for the largest simulation box \( L_\mathrm{0}\) = 600, 200 points are no longer randomly picked but regularly arrayed.}
  \label{f1}
\end{figure}

Next, all melt systems are quenched to $T$ = 0.77 (423.5 K) for isothermal crystallization, during which the time step is chosen as 0.01 ($\sim$ 35 fs), and the periodic boundary conditions are applied on all three axes. 
The crystallization proceeds in the NPT ensemble with the pressure $P$ = 8 (1 atm), where the pressure and temperature are controlled by the Nosé-Hoover barostat and thermostat with a damp time of 1000 and 100 MD time steps, respectively. The centrosymmetry parameter (\( P_\mathrm{cs} \)) \cite{ref33} and the length of successive trans-trans monomers (\( L_\mathrm{tt} \)) \cite{ref34} are applied to identify the crystalline monomers. The nucleation criterion, i.e., \( P_\mathrm{cs} < 1 \) and \( L_\mathrm{tt} \geq 8 \) is determined according to the distribution analyses of \( P_\mathrm{cs} \) and \( L_\mathrm{tt} \), which is included in S2 and Figure S2 of the SI. 

It should be noticed that systems with \( \left \langle  N_{\mathrm{e0}}\right \rangle\) = 6 $\sim$ 15 are also built. However, the nucleation of these systems is extremely difficult and takes a long incubation period. In addition, disentanglement takes place during the incubation, leading to a significant deviation of the entanglement densities from the initial set values when the nucleation occurs. Therefore, these systems are not included in the current work.

\subsection{2.2. Analysis of the nucleation free energy barrier}

The nucleation barrier was obtained by using the mean first-passage time (MFPT) method proposed by Wedekind et al \cite{ref32}. The MFPT method provides a simple and efficient strategy to analyze MD trajectories of activated processes. By using the MFPT method, the information of the nucleation process can be directly extracted from the trajectories of MD simulations, including the nucleation free energy barrier and the critical nucleus size. However, the huge difference between the time scales of nucleation and growth would affect the analysis results. 
To overcome this problem, Nicholson et al. introduced an analysis method based on the MFPT, and applied it to the simulation of nucleation in $n$-eicosane successfully \cite{ref35,ref36}. This modified MFPT method allows us to extract valuable nucleation kinetic parameters from the simulation data of the nucleation process. The detailed fitting process can be found in S4 of the SI. Then, we can attempt to understand the evolution trend of the nucleation barrier with the entanglement density in the framework of the CNT.

The CNT is a simple and flexible framework to describe nucleation phenomena. Based on the CNT, the nucleation free energy barrier of a cylindrical nucleus is given as \cite{ref37,ref38}:
\begin{equation}
  \Delta G_\mathrm{CNT}^{*} = 2 \sigma_l^*\sigma_f/\varepsilon^2 
  \label{4}
\end{equation}

where \( \sigma_l\) and \( \sigma_f\) are the lateral and fold surface free energy per monomer, respectively. $\varepsilon$ is the bulk free energy of the formation of a nucleus per monomer:
\begin{equation}
  \varepsilon=\Delta H_i - T(\Delta S_{\mathrm{con},i} + \Delta S_\mathrm{atom} )
  \label{5}
\end{equation}

where \(\Delta H_i\)  represents the enthalpy change of a monomer, \( \Delta S_{\mathrm{con},i} \) and \( \Delta S_\mathrm{atom} \) are the local conformational entropy and atomic entropy changes of a monomer during the nucleation, respectively. Here we are interested in the variation trend of \(\Delta G^{*}\) with \( \left \langle  N_{\mathrm{e0}}\right \rangle\) rather than the absolute value of \(\Delta G^*\). Thus, based on the simulation results, the variation trends of each term in the right side of Eq. (4) with \( \left \langle  N_{\mathrm{e0}}\right \rangle\) are evaluated independently in the following sections. (See S5-7 in the SI for the calculation details) 

\section{3. Results and theory}

\subsection{3.1. Simulation results}

The polymer melt systems with different initial entanglement densities show significantly different nucleation behaviors. Figures 2a and 2b present snapshots of the nucleus formation and early growth of the systems with \( \left \langle  N_{\mathrm{e0}}\right \rangle\) = 267 and \( \left \langle  N_{\mathrm{e0}}\right \rangle\) = 17, respectively. Nucleation of the former system occurs rapidly involving just a few polymer chains. In contrast, nucleation of the latter system takes a long incubation time involving multiple polymer chains. It is found that the higher entanglement density is, the slower nucleation rate becomes, which is consistent with previous experimental and simulation results \cite{ref21,ref39,ref40}.

\begin{figure}
  \centering
  \includegraphics[width=12cm,height=7.28cm]{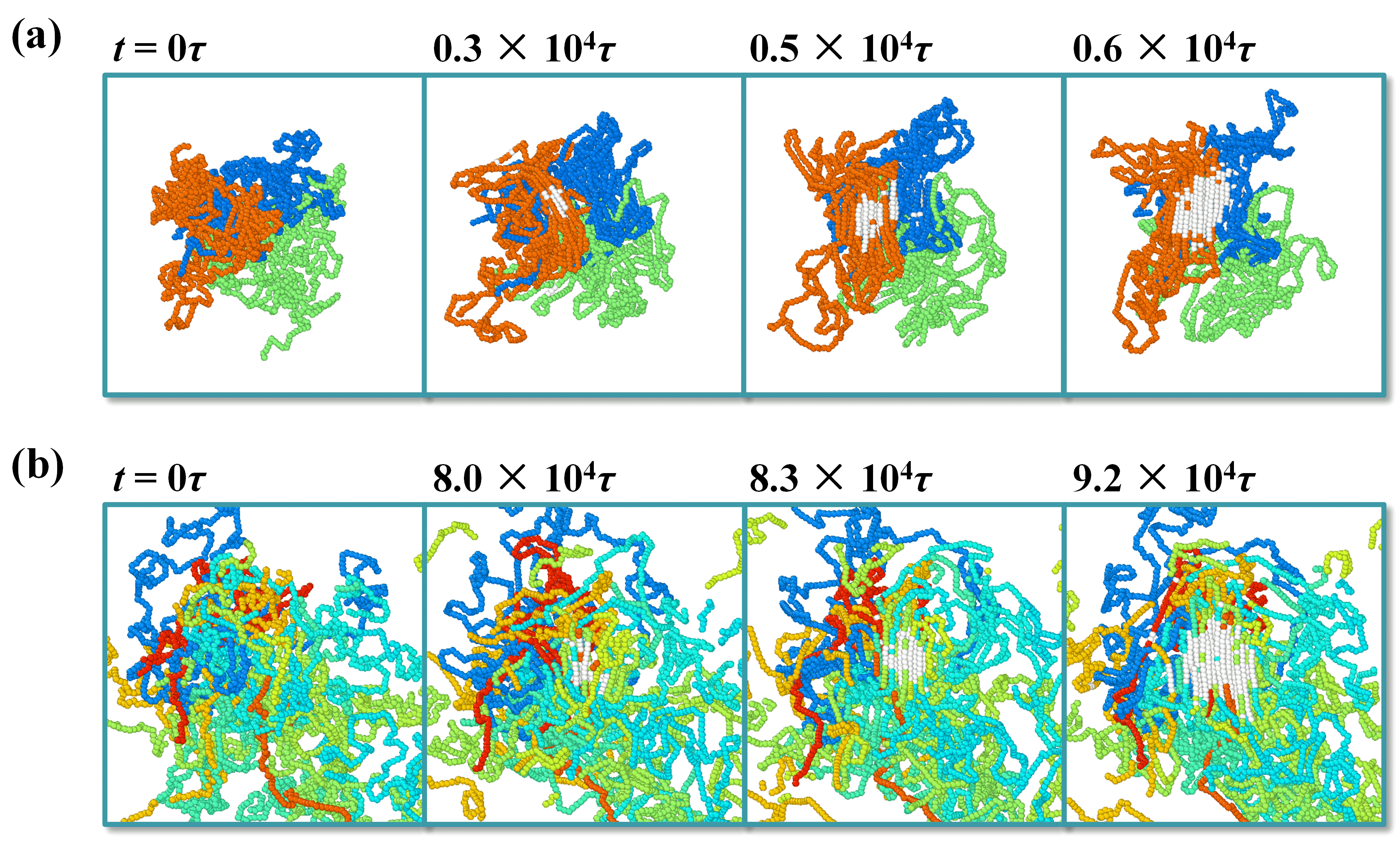}
  \caption{Snapshots of the formation and growth of the primary nucleus in systems with (a) \( \left \langle  N_{\mathrm{e0}}\right \rangle\) = 267, and (b) \( \left \langle  N_{\mathrm{e0}}\right \rangle\) = 17. Only chains involved in the formation of a nucleus are drawn here. Segments in the amorphous part and the nucleus are marked in different colors and white, respectively.}
  \label{f2}
\end{figure}

The effect of entanglement on the nucleation kinetics is evaluated by calculating the mean square displacement (MSD) of monomers and the center of mass of a chain.  The expressions for the MSD of monomers and the center of mass of a chain are \(g_1(t) = \left \langle [r_i(t) - r_i(0)]^2\right \rangle \), 
and \(g_3(t) = \left \langle [r_{\mathrm{cm}}(t) - r_{\mathrm{cm}}(0)]^2\right \rangle \), respectively.
Here, \( r_i\) and \( r_\mathrm{cm}\)  are the coordinates of the $i$th monomer and the center of mass of a chain, respectively. Figure 3a presents the variations of \(g_1\) and \(g_3\) with time for all polymer melt systems. The curves of \(g_1(t)\) for all systems overlap almost completely, suggesting that the movement of monomers is not sensitive to the entanglement density. 
\(g_3(t)\) characterizes the movement of a polymer chain, and is found to be independent of the entanglement density at short time scales, while becomes diverging at long time scales. Namely, the diffusion of a polymer chain at the nucleus interface during crystal growth is restricted by the entanglement. In brief, although the diffusion of a polymer chain scale is indeed influenced by the entanglement, its effect on the diffusion of monomers is quite limited. Considering the nucleation is a behavior occurring at the monomeric scale, the diffusion at the polymer chain scale should not be responsible for the significantly different nucleation rates of systems with different entanglement densities.

\begin{figure}
  \centering
  \includegraphics{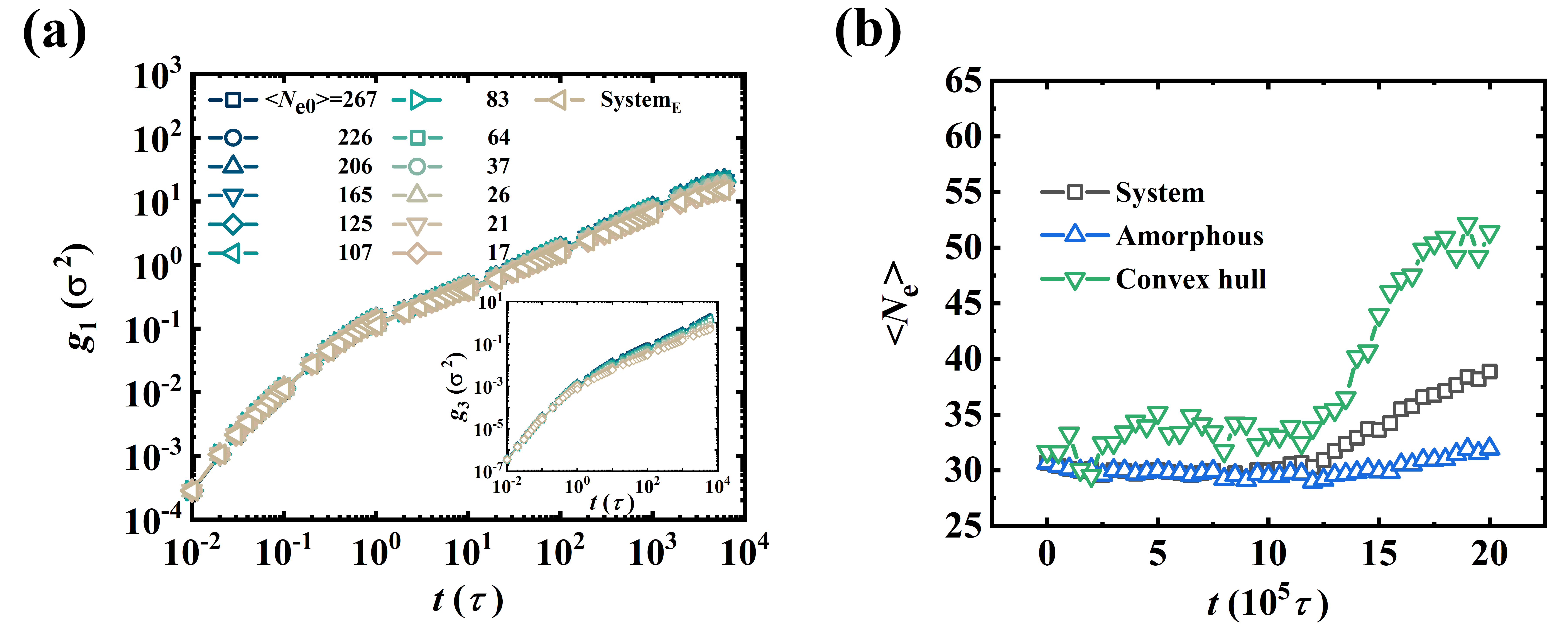}
  \caption{(a) The MSD of monomers (\(g_1\)) for all polymer melt systems. The inset is the MSD of the center of mass of polymer chains (\(g_3\)) for all systems. (b) The change of average entanglement length in the whole system, the amorphous region, and the convex hulls during the crystallization process.}
  \label{f3}
\end{figure}

To determine whether the long nucleation incubation period is related to the disentanglement process, the evolution of \( \left \langle  N_{\mathrm{e}}\right \rangle\) for the system with \( \left \langle  N_{\mathrm{e0}}\right \rangle\) = 17 during the nucleation is measured. First, the spatial positions of critical nuclei are determined. The three-dimensional region surrounded by the boundary of each nucleus is defined as the convex hull. The value of the entanglement length \( N_{\mathrm{e}}\) is assigned to each monomer in the corresponding entangled segment. The obtained ensemble average entanglement lengths \( \left \langle  N_{\mathrm{e}}\right \rangle\) in the whole system, the amorphous region, and the convex hulls are shown in Figure 3b. In this system, most nuclei reach the critical nucleus size and start to grow at 12 $\sim$ 15 $\times$ \(10^5 \tau\). As can be seen, \( \left \langle  N_{\mathrm{e}}\right \rangle\) in the convex hulls keeps almost constant, indicating that no obvious disentanglement occurs during the nucleation. It is generally considered that the nucleation rate is determined together by the activation energy for the transport process at the interface between the melt and the nucleus surface (\(\Delta E\)) and the nucleation barrier (\(\Delta G^*\)), i.e., \(J = J_0exp[-\frac{\Delta E}{RT}\,-\frac{\Delta G^*}{RT}\,]\) \cite{ref41}. The above calculations demonstrate clearly that the effect of the entanglement density on the nucleation rate mainly stems from changing \(\Delta G^*\) rather than \(\Delta E\). 

By applying the modified mean first-passage time (MFPT) method \cite{ref35} (See S4, Figure S3, and Table S2 of the SI for calculation details) to 50 MD trajectories of each polymer system, we can obtain the nucleation free energy barrier (\(\Delta G^*\)) and critical nucleus size (\(n^*\)), and plot them in Figures 4a and 4b. With the increase of \( \left \langle  N_{\mathrm{e0}}\right \rangle\), both \(\Delta G^*\) and \(n^*\) decrease, and the downtrends slow down as \( \left \langle  N_{\mathrm{e0}}\right \rangle > 64\) . \(\Delta G^*\) and \(n^*\) show an approximately linear relationship, which is in agreement with the CNT if the bulk free energy density $\varepsilon$ for different systems is the same. As shown in Figure S4 of the SI, an almost linear relationship between \(\Delta G^*\) and 1/\( \left \langle  N_{\mathrm{e0}}\right \rangle\) (number density of entanglement) is found, which is consistent with the previous experimental study \cite{ref21}.

\begin{figure}
  \centering
  \includegraphics{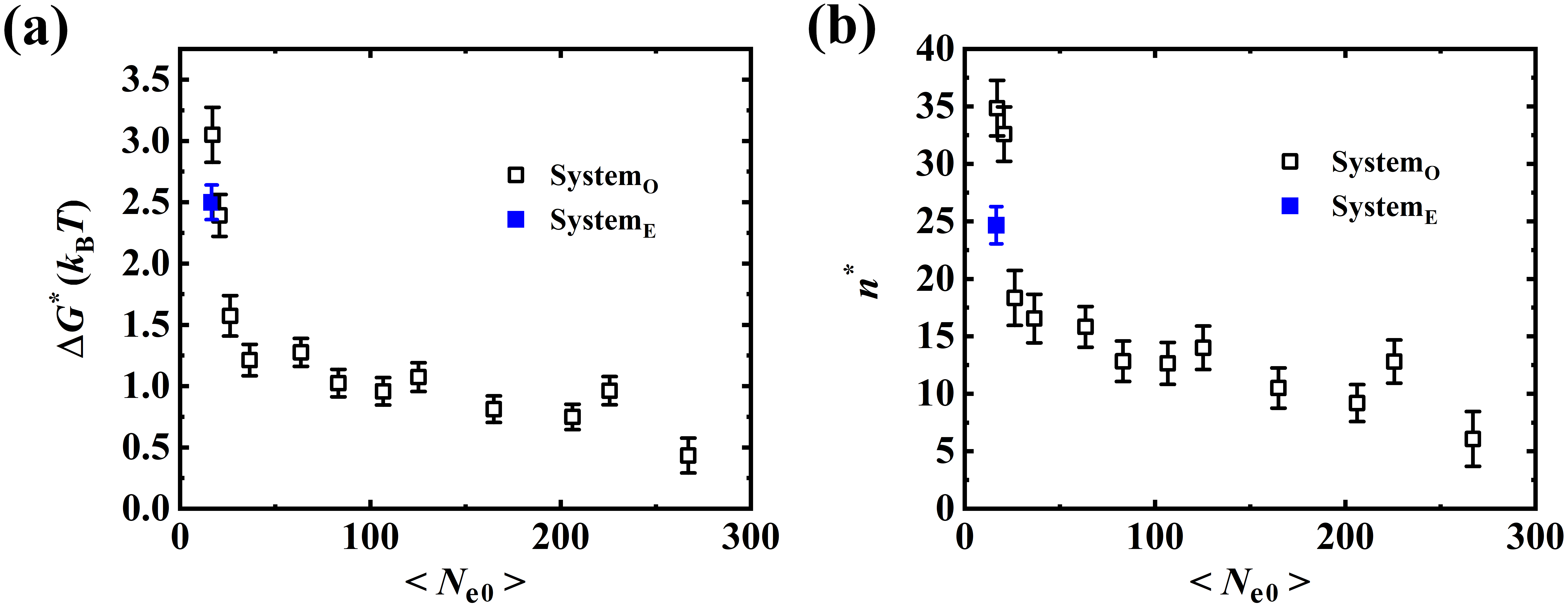}
  \caption{(a) The nucleation free energy barrier \(\Delta G^*\), and (b) the critical nucleus size \(n^*\) as a function of the entanglement length \( \left \langle  N_{\mathrm{e0}}\right \rangle\). The solid blue and open-black symbols represent the systems in and out of equilibrium, respectively.}
  \label{f4}
\end{figure}

According to Eq. (4), to verify whether the CNT can account for the nucleation barrier presented in Figure 4a, we need to figure out how the lateral surface free energy per monomer \(\sigma_l\), the fold surface free energy per monomer \(\sigma_f\), and the bulk free energy of the formation of a nucleus per monomer $\varepsilon$ vary with the entanglement length \( \left \langle  N_{\mathrm{e0}}\right \rangle\). 
Our first concern here is $\varepsilon$. Specifically,
the enthalpy change per monomer \( \Delta H_i = \Delta H/N_\mathrm{c}\) keeps constant for polymer systems with different \( \left \langle  N_{\mathrm{e0}}\right \rangle\),
and is estimated to be \(-0.77k_\mathrm{B}T\), where \(\Delta H\) is the total enthalpy loss, and \(N_\mathrm{c}\) is the number of monomers in the nucleus.
The conformational entropy change \(\Delta S_\mathrm{con}\) is calculated from the end-to-end distance distribution of the chains \cite{ref42} (see S5 and Figure S5 of the SI for the calculation details). As shown in Figure 5a, \(\Delta S_\mathrm{con}\) decreases slightly due to the relaxation of the systems and the increased chain stiffness caused by quenching.
\(\Delta S_{\mathrm{con},i} = \Delta S_\mathrm{con}/N_\mathrm{t}\) with \(N_\mathrm{t}\) being the total number of monomers 2 $\times$ \(10^5\) is rather small and thus can be negligible during the nucleation. The atomic entropy \(S_\mathrm{atom}\) describes the periodic symmetry property of the system, and can be used to characterize the configurational entropy of a nucleus. 
Based on the quadratic term in the expansion of the excess entropy of liquid, the configurational entropic difference \(\Delta S_\mathrm{atom}\) of a monomer between the target system and the ideal gas can be calculated \cite{ref43,ref44,ref45}. 
Figure 5b suggests that \(\Delta S_\mathrm{atom}\) hardly changes with the entanglement density. It should be pointed out that the obtained data is a relative value owing to the used Dirac broadening function and local averaging (see S6 and Figure S6 in the SI for details). 
Furthermore, a prefactor \(\beta_1\) is required for numerical corrections. Hereto, we can conclude that $\varepsilon$ does not vary with the entanglement density, and thus is not be responsible for the decreasing \(\Delta G^*\) with the increase of \( \left \langle  N_{\mathrm{e0}}\right \rangle\).

\begin{figure}
  \centering
  \includegraphics{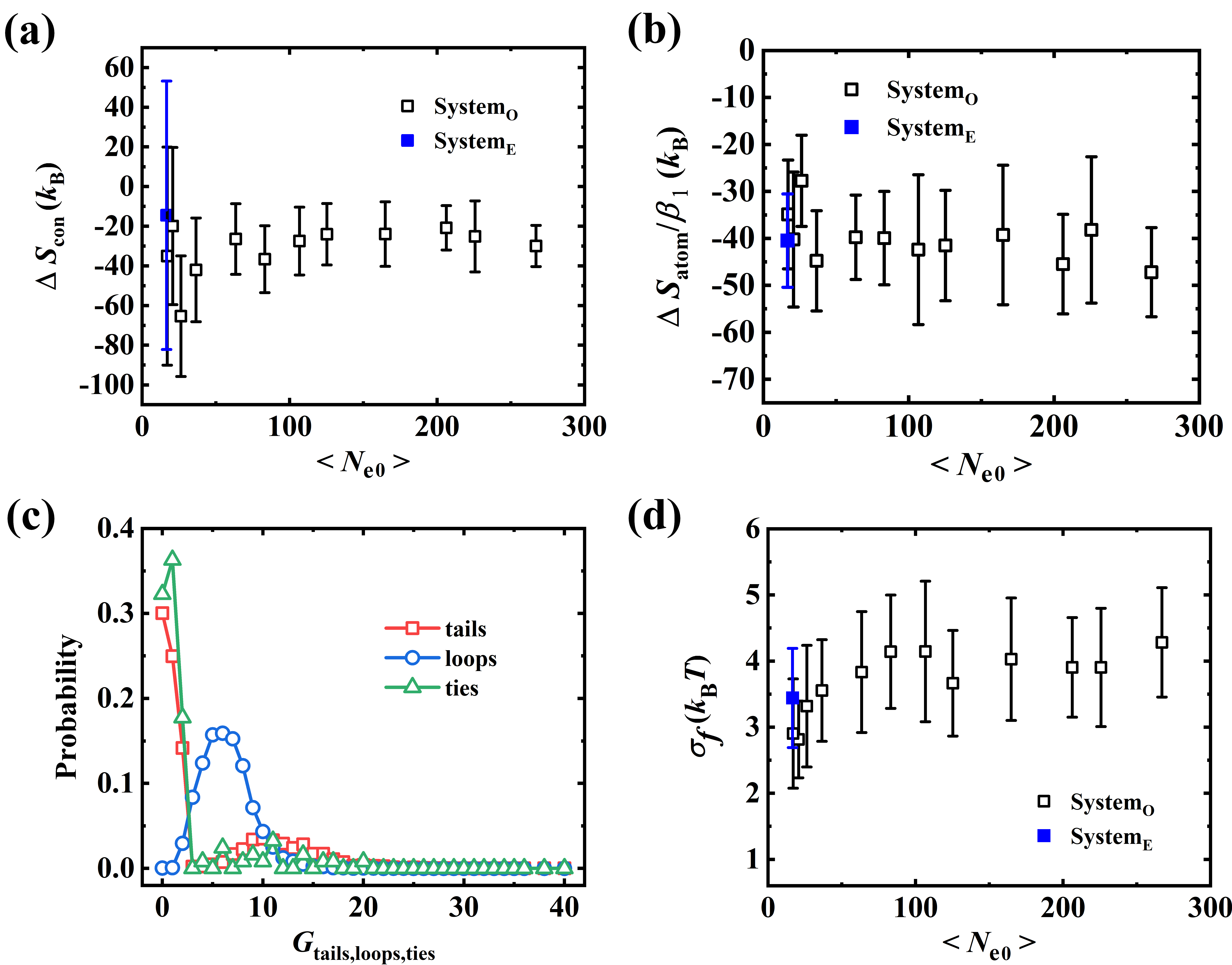}
  \caption{Variations of (a) the conformational entropy of polymer chains, and (b) the atomic entropy of polymer systems with different entanglement densities during nucleation. (c) The distributions of the free energy of the amorphous structures (tails, loops, and ties). (d) The fold surface free energy density of a nucleus as a function of the entanglement length. All data points are obtained by averaging the calculation results of 10 MD trajectories of each system.}
  \label{f5}
\end{figure}

As to the lateral surface free energy \(\sigma_l\), it is mainly determined by the interaction potential difference between a nucleus and its surrounding amorphous. Thus, \(\sigma_l\) should be dependent on the lattice structure and the amorphous density, and is irrelevant to the entangled state of the polymer systems. In other words, \(\sigma_l\) can be considered as a constant independent of \( \left \langle  N_{\mathrm{e0}}\right \rangle\). \(\sigma_f\) is determined by the entropy loss in the amorphous region during the nucleation rather than by the potential difference \cite{ref24}. After extracting nuclei from the polymer systems, the free energy of the amorphous structures (tails/loops/ties) on the fold surface is calculated according to the theory of Muthukumar \cite{ref46} (see S7 of the SI for details). Figure 5c shows the distributions of the free energy of amorphous structures (\(G_\mathrm{tails,loops,ties}\)). The free energy of loops \(G_\mathrm{loops}\) is higher than those of tails \(G_\mathrm{tails}\) and ties \(G_\mathrm{ties}\), which can be attributed to the lower conformational entropy of loops. Figure S7 shows that the formation of loops prevails over that of tails and ties, giving rise to an increase of \(\sigma_f\) with the increasing \( \left \langle  N_{\mathrm{e0}}\right \rangle\) shown in Figure 5d. The above analyses show clearly that in the framework of the CNT, the nucleation free energy barrier rises as the entanglement length increases. Obviously, this prediction does not agree with our simulation results. Actually, the failure of the CNT in describing the nucleation of polymer systems is not surprising since it does not take into account the unique connectivity of polymer chains, and the accompanying effects such as the entanglement. 

\subsection{3.2. The entanglement free energy theory}
During the nucleation, the freedom of an amorphous segment gets reduced due to its transformation into a crystal stem. Considering that the constraint imposed by the crystal stem is similar to the entanglement, the crystal stem can be regarded as an additional entanglement point \cite{ref47}. The entanglement state of the amorphous segments is changed by the formation of crystal stems, which might play an important role in the nucleation and cause a variation of the nucleation barrier with the entanglement density. To capture the underlying physics of the nucleation in polymer systems, we propose a theoretical approach in which the entanglement free energy (\(G_\mathrm{z}\)) is introduced to quantify the effect of entanglement on nucleation. \(G_\mathrm{z}\) originates from the redistribution of the entanglement length caused by the slip of entanglement points, and can be expressed as \(G_\mathrm{z} = U_\mathrm{z} - TS_\mathrm{z}\). Here, the entanglement energy (\(U_\mathrm{z}\)) describes the repulsive interactions between entanglement points, and the entanglement entropy (\(S_\mathrm{z}\)) reflects the entanglement density and its spatial distribution, which compensates the attractive interactions between two adjacent entanglement points. 

The slip-link model predicts that systems with entanglement densities below and above the equilibrium value have the same free energy \cite{ref47,ref48,ref49,ref50}. Clearly, it cannot explain the increased nucleation barrier as the entanglement density rises. Inspired by the slip-link model and also the local-knots model \cite{ref51,ref52}, we use the harmonic oscillator model to describe the motion of entanglement points. In the harmonic oscillator model, adjacent entanglement points in a polymer chain are considered to interact repulsively, and the repulsive strength is determined by the distance between the adjacent entanglement points (entanglement length \(N_\mathrm{e}\)). Under the thermal fluctuations, entanglement points keep sliding along the chain, and the global entanglement network changes dynamically. Note that this change induced by thermal fluctuations is slight when the observation period is short or the entanglement state is stable. In these cases, an entanglement point just oscillates nearby its original position. Similar to the slip-link model, a polymer chain can be considered as a series of untangled segments connected by small springs (i.e., entanglement points).

Let’s suppose that the entanglement points are uniformly distributed in a polymer chain, and each entanglement point can be treated as an independent harmonic oscillator to describe its local slip. Then, the one-dimensional motion of an entanglement point is controlled by a harmonic potential:
\begin{equation}
  U_\mathrm{z} = \frac{1}{2}\,k_\mathrm{s} \Delta x^2
  \label{6}
\end{equation}
where \(U_\mathrm{z}\) is the entanglement energy, which is the energy of each entanglement point. \(\Delta x\) is the slip distance of an entanglement point under thermal fluctuations, which is in the units of a coarse-grained bead. \(\Delta x = N_\mathrm{e}\) in the limit case that two adjacent entanglement points overlap. Polymer chains with $N$ monomers in each chain are involved in the entanglement overlapping, during which the freedom of monomers decreases. We take monomer number $N$ as the scale parameter and the energy penalty for the overlapping is assumed to be \(\gamma Nk_\mathrm{B}T\) with $\gamma$  being a constant. Here, \(Nk_\mathrm{B}T\) can be considered as the reduced unit. Then, the stiffness coefficient of the oscillator can be deduced as \(k_\mathrm{s} = 2\gamma Nk_\mathrm{B}T/N_\mathrm{e}^2\). Obviously, \(k_\mathrm{s}\)  is a variable relying on the chain length $N$ and the entanglement length \(N_\mathrm{e}\). Thus, we have:
\begin{equation}
  U_\mathrm{z} = \frac{1}{2}\,k_\mathrm{s} \Delta x^2 \xrightarrow{\beta_2 = \gamma \Delta x^2} \beta_2Nk_\mathrm{B}T/N_\mathrm{e}^2
  \label{7}
\end{equation}
Here, \(\beta_2 = \gamma \Delta x^2\) is the prefactor. During the nucleation, the entanglement network is considered to be stable, and its evolution occurs at a small spatial scale. It implies that the value of \(\Delta x\) is small and keeps nearly constant. Then, \(\beta_2\) can be approximately treated as a constant for polymer systems with different \(N_{\mathrm{e}}\). Most importantly, Eq. (7) predicts the relationship between the entanglement energy and the entanglement length, i.e., \(U_\mathrm{z} \sim 1/N_\mathrm{e}^2 \). A higher \(U_\mathrm{z}\) with smaller \(N_\mathrm{e}\) implies that a more untangled state is preferred (\(N_\mathrm{e}\) increases). Note that a dramatic variation in \(U_\mathrm{z}\) does not originate from the vibration of the harmonic oscillator itself (\(\Delta x\)), but a sharp change in \(N_\mathrm{e}\) induced by the formation of crystal stem.

The change of the entanglement state of the amorphous segments in nucleation is accompanied by a variation in \(U_\mathrm{z}\). However, the crystal stem is considered to be fixed (\(\Delta x = 0\)), and does not contribute to the variation in \(U_\mathrm{z}\). The formation of a crystal stem also keeps the adjacent entanglement points away. Namely, the constraints brought by the crystal stem do not affect the original entanglement network, but can restrict its evolution. Consider the case where a crystal stem forms at the midpoint of the amorphous segment of the length \(N_\mathrm{e}\) during the time interval \([t_1,t_2]\). As a result, two new amorphous segments of the length \(N_\mathrm{e}/2\) form. Since the nucleation takes place at the time scale much smaller than the reputation time, the assumption that \(\Delta x = 1\) during the nucleation process is reasonable. Therefore, the entanglement energy change due to the formation of a crystal stem can be given as:
\begin{equation}
  \Delta U_\mathrm{z} = U_{\mathrm{z},t_2} - U_{\mathrm{z},t_1} = \beta_26Nk_\mathrm{B}T/ N_\mathrm{e}^2
  \label{8}
\end{equation}
Here, \(U_{\mathrm{z},t_1} = \beta_22Nk_\mathrm{B}T/ N_\mathrm{e}^2\) is the entanglement energy of the amorphous segments between two adjacent entanglement points at the time \(t_1\) with \( N_\mathrm{e}(t=t_1) = N_\mathrm{e}\). When a crystal stem at the time \(t_2\), \(U_{\mathrm{z},t_2} = \beta_28Nk_\mathrm{B}T/ N_\mathrm{e}^2\) with \( N_\mathrm{e}(t=t_2) = N_\mathrm{e}/2\).

Next, let’s turn our attention to the entanglement entropy \(S_\mathrm{z}\). The concept of \(S_\mathrm{z}\) comes from the excess entropy. By expanding the configurational entropy of simple liquids with the many-body correlation function, the second term only involving the correlation function is usually called the two-body excess entropy, which makes a nearly 90$\%$ contribution to the configurational entropy \cite{ref43,ref53}. Polymer chains are divided into multiple untangled segments by the entanglement points. The length of these untangled segments is the entanglement length \(N_\mathrm{e}\). The distribution of \(N_\mathrm{e}\) is ever-changing due to the slip of entanglement points. Essentially, \(N_\mathrm{e}\) is a reflection of the interaction between two adjacent entanglement points, i.e., a special two-body interaction. This two-body interaction can be well described by the excess entropy, and the magnitude of excess entropy is a function of the number of entanglement points. 

The original excess entropy is the configurational entropy, which incorporates the interactions between any two particles in the system. Considering our focus here is to describe the two-body interactions between entanglement points, the probability density of the entanglement length \(N_\mathrm{e}\) rather than the radial distribution of entanglement points in the expression of the original excess entropy (\(f(r)\) in Eq. S8.1) is used to calculate \(S_\mathrm{z}\). For the sake of simplicity, the calculations are carried out in the spherical coordinate system, which is referred to as the $Z$-space in this work. As shown in Figure 6, an untangled segment of the length \(N_\mathrm{e}\) between two adjacent entanglement points along a chain in the real space is projected as a point with a distance \(r = N_\mathrm{e}\) from the origin in the $Z$-space. Here, the angular coordinates of the points are ignored. The \(N_\mathrm{e}\) calculated by the Z1 code is taken as an integer value, so $r$ in the $Z$-space takes discrete values like 1, 2, ..., $N$. In this way, the probability density function of \(N_\mathrm{e}\) in the real space is converted into that of the distance r in the $Z$-space, \(g(r)\). 

For an ideally untangled system consisting of \(N_\mathrm{chain}\) chains of the length $N$, $r$ = $N$. That is, there are \(N_\mathrm{chain}\) points with $r$ = $N$ in the $Z$-space. Thus, we should have \(g_0(r = N) = 1\). \(g(r)\) would be different for a given entangled system. Then, \(S_\mathrm{z}\) is defined as the entropic difference of points in the $Z$-space projected by the given entangled systems and the ideally untangled system (see S8 of SI for the detailed derivation process):
\begin{equation}
  S_\mathrm{z} = \beta_3k_\mathrm{B}\int_1^N[g(r)lng(r)-g_0(r)lng_0(r)-g(r)+g_0(r)]r^2\mathrm{d}r 
  \label{9}
\end{equation}
where $r$ is the distance of a point from the origin in the $Z$-space, and \(\beta_3\) is a constant dependent on the system size and chain length. Since the density of points in the $Z$-space is independent of the shell volume, the probability density function rather than the radial distribution function is used here. In this way, the normalization of the integral volume is not necessary. Moreover, an additional point is added to each $r$ in the $Z$-space to fill the singularity in the integral function. The influence of the additional points is negligible for sufficiently long segments as the probability density function of the additional points \(g(r) = 1/N\) is small.

\begin{figure}
  \centering
    \includegraphics[width=12cm,height=5.6cm]{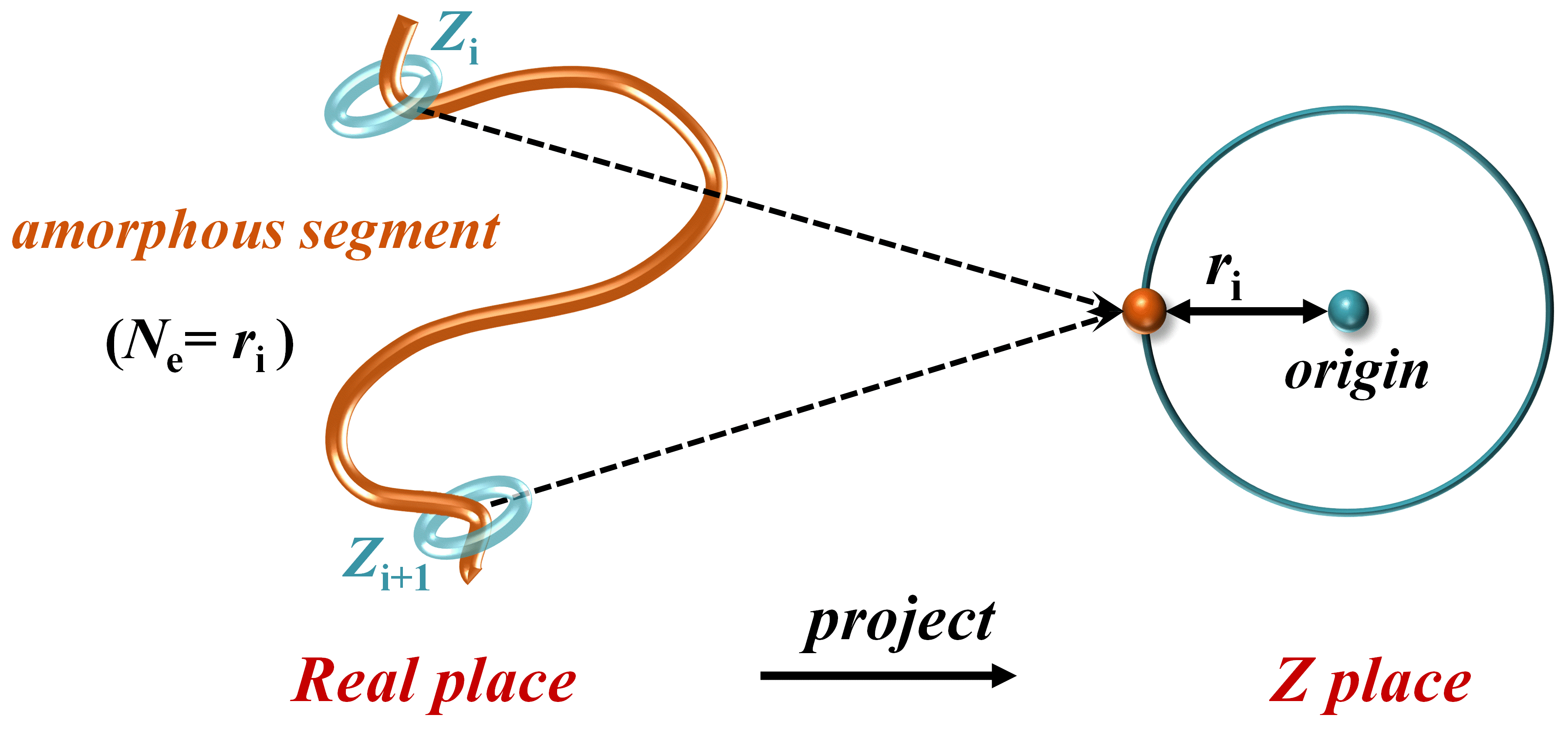}
  \caption{Schematic illustration of the projection of two adjacent entanglement points in the real space into the $Z$-space. \(N_\mathrm{e}\) is the entanglement length between two adjacent entanglement points.}
  \label{f6}
\end{figure}

\begin{figure}
  \centering
  \includegraphics{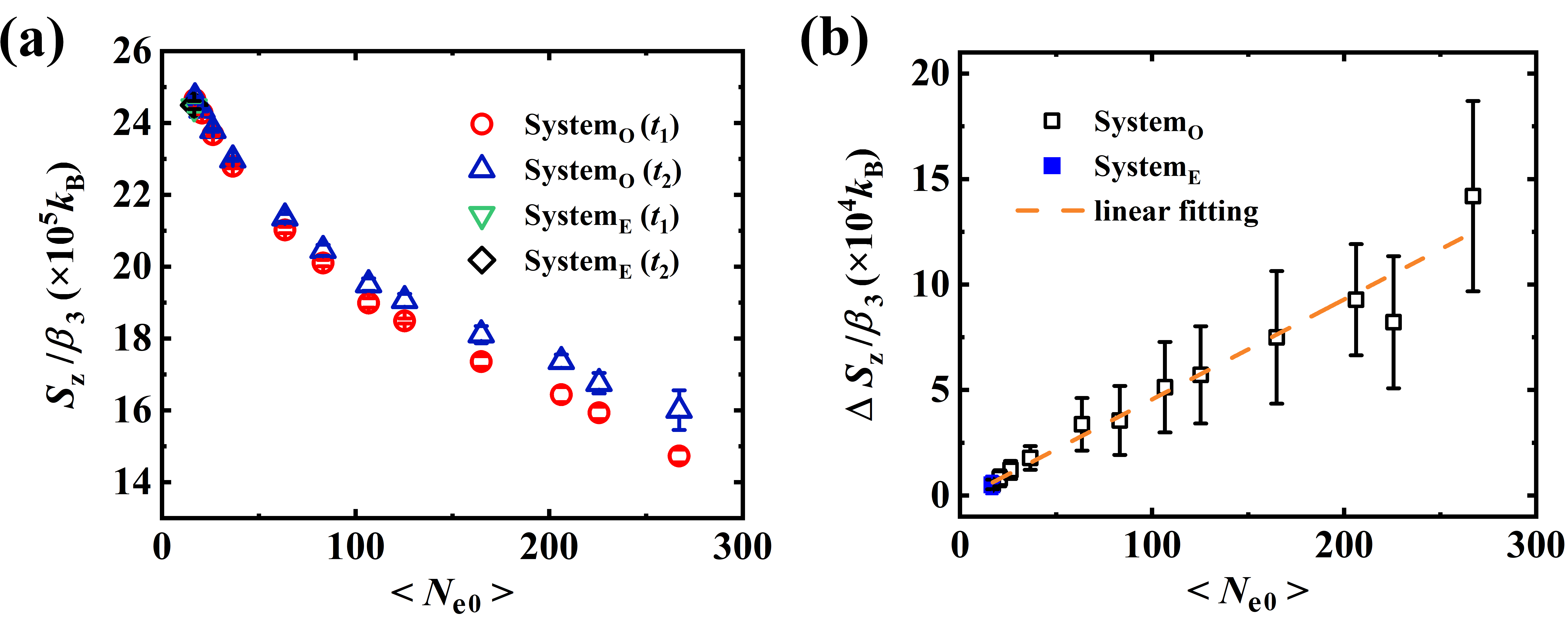}
  \caption{(a) \(S_\mathrm{z}\) of the whole systems at the beginning and the ending of nucleation. (b) The variations in \(S_\mathrm{z}\) of the whole system during the nucleation with the entanglement density.}
  \label{f7}
\end{figure}

Evolutions of \(S_\mathrm{z}\) of systems at the beginning and the ending of nucleation with the entanglement density are plotted in Figure 7a. \(S_\mathrm{z} = 0\) for untangled systems, and \(S_\mathrm{z} > 0\) for entangled systems. Again, consider the case where a portion of an amorphous segment of the length \(N_\mathrm{e}\) turns into a crystal stem during the time interval \([t_1,t_2]\). Then, the variation of \(S_\mathrm{z}\) of this segment due to the formation of the crystal stem is:
\begin{equation}
  \Delta S_\mathrm{z} = \beta_3k_\mathrm{B}\int_1^{N_\mathrm{e}} [g_{t_2}(r)lng_{t_2}(r)-g_{t_1}(r)lng_{t_1}(r)-g_{t_2}(r)+g_{t_1}(r)]r^2\mathrm{d}r 
  \label{10}
\end{equation}
where \(g_{t_1}(r)\) and \(g_{t_2}(r)\) are the probability density functions of points in the $Z$-space at \(t = t_1\) and \(t_2\), respectively. Generally, the length of a crystal stem in the primary nucleus is much smaller than \(N_\mathrm{e}\). Assuming that the nucleation occurs at the midpoint of the segment and the crystal stem is treated as an additional entanglement point, the segment can be divided into two subsegments of the length ca. \(N_\mathrm{e}/2\) at the time \(t_2\). In the $Z$-space, there is a point at \(r = N_\mathrm{e}\) and two points at \(r = N_\mathrm{e}/2\). A virtual point is added at each $r$ in \(r = [1,N_\mathrm{e}]\) to compensate singularities. Then, the total number of points in the $Z$-space at \(t = t_1\) and \(t = t_2\) is \(N_\mathrm{e} + 1\) and \(N_\mathrm{e} + 2\), respectively. Therefore, the probability density function in $Z$-space can be given as:
\begin{equation}
g_{t_1}(r) = 
\left\{   
    \begin{array}{lr}
         \displaystyle{\frac{2}{N_\mathrm{e} + 1}\,,r = N_\mathrm{e},} & \\
         &\\
         \displaystyle{\frac{1}{N_\mathrm{e} + 1}\,,r \neq N_\mathrm{e}.} & \\
    \end{array}
\right.
\label{11}
\end{equation}

\begin{equation}
g_{t_2}(r) = 
\left\{   
    \begin{array}{lr}
         \displaystyle{\frac{3}{N_\mathrm{e} + 2}\,,r = N_\mathrm{e}/2,} & \\
         &\\
         \displaystyle{\frac{1}{N_\mathrm{e} + 2}\,,r \neq N_\mathrm{e}/2.} & \\
    \end{array}
\right.
\label{12}
\end{equation}
Here, $r$ ranges from 1 to $N$. For a sufficiently long amorphous segment with \(N_\mathrm{e} \gg 1\), the denominators in Eqs. (11) and (12) can be approximated as \(N_\mathrm{e}\). Substituting Eq. (11) and Eq. (12) into Eq. (10), we get:
\begin{equation}
  \Delta S_\mathrm{z} = \beta_3k_\mathrm{B}N_\mathrm{e}\left[\frac{1}{2}\left(lnN_\mathrm{e}+1\right)+\frac{3}{4}ln3-2ln2\right] \approx \beta_3k_\mathrm{B}N_\mathrm{e}\left(\frac{1}{2}lnN_\mathrm{e}-0.0623\right)
  \label{13}
\end{equation}
The detailed derivation of Eq. (13) can be found in S9 of the SI. Eq. (13) suggests that the entanglement entropy gained from the formation of a crystal stem is approximately proportional to \(N_\mathrm{e}\), which is consistent with our simulation results shown in Figure 7b.

Combining Eqs. (8) and (13), we obtain the change in the entanglement free energy change during nucleation:
\begin{equation}
  \Delta G_\mathrm{z} = \mu(\Delta U_\mathrm{z} - T\Delta S_\mathrm{z}) \approx \mu k_\mathrm{B}T \left[\beta_2\frac{6N}{N_\mathrm{e}^2} - \beta_3N_\mathrm{e}	\left(\frac{1}{2}lnN_\mathrm{e} - 0.0623\right)\right]
  \label{14}
\end{equation}
where $\mu$ is the number of crystal stems inside a nucleus. Then, the total change in the free energy for the formation of a nucleus with $\mu$ stems of the length \(l_\mathrm{s}\) is given as:
\begin{equation}
  \Delta G = \Delta G_\mathrm{CNT} + \Delta G_\mathrm{z} = -\mu l_\mathrm{s}\varepsilon + \sqrt{\mu}l_\mathrm{s}\sigma_{l} +2\mu\sigma_{f} + \Delta G_\mathrm{z}
  \label{15}
\end{equation}
Minimizing \(\Delta G\) with respect to $\mu$ and \(l_\mathrm{s}\), the free energy barrier for the nucleation is:
\begin{equation}
  \Delta G^* = 2\sigma_l^2\Lambda/\epsilon^2
  \label{16}
\end{equation}
where
\begin{equation}
  \Lambda = \sigma_f + \frac{1}{2}k_\mathrm{B}T\left[\beta_2\frac{6N}{N_\mathrm{e}^2} - \beta_3N_\mathrm{e}\left(\frac{1}{2}\,lnN_\mathrm{e}-0.0623\right) \right]
  \label{17}
\end{equation}

Note that the nucleation barrier given by Eq. (16) has the same mathematical form as that given by the CNT (Eq. (4)). $\Lambda$ consists of two parts, \(\sigma_f\) in Eq. (17) inherits from the CNT. The second term is the correction contributing from the entanglement free energy, characterizing the disturbance on the entanglement network around the nucleus by the nucleation. Since the entanglement state of each system when nucleation occurs is similar to the initial entanglement, the entanglement length \(N_\mathrm{e}\) is approximate to its initial value \( \left \langle  N_{\mathrm{e0}}\right \rangle\). With the already obtained \(\sigma_f\) (Figure 5d), $\varepsilon$ and \(N_\mathrm{e} \approx \left \langle  N_{\mathrm{e0}}\right \rangle\), the validity of Eq. (16) can be verified. As shown in Figure 8a, the fitted values of \(\Delta G^*\) agree well with these obtained by the MFPT method (Figure 4a). The used parameters in the fittings are \(\beta_1 = 4.28 \times 10^{-4}\), \(\beta_2 = 0.356\) and \(\beta_3 = 0.0027\), respectively. \(\beta_1\) is the prefactor of atomic entropy in the bulk free energy term $\varepsilon$. Note that the difference in the magnitude of \(\beta_2\) and \(\beta_3\) is due to the magnitudes of \(6N/N_\mathrm{e}^2\) and \(\beta_3N_\mathrm{e}\left(\frac{1}{2}\,lnN_\mathrm{e}-0.0623\right)\) are 1 and 100, respectively. The energy and the entropy contributions are of the same order.

\begin{figure}
  \centering
  \includegraphics{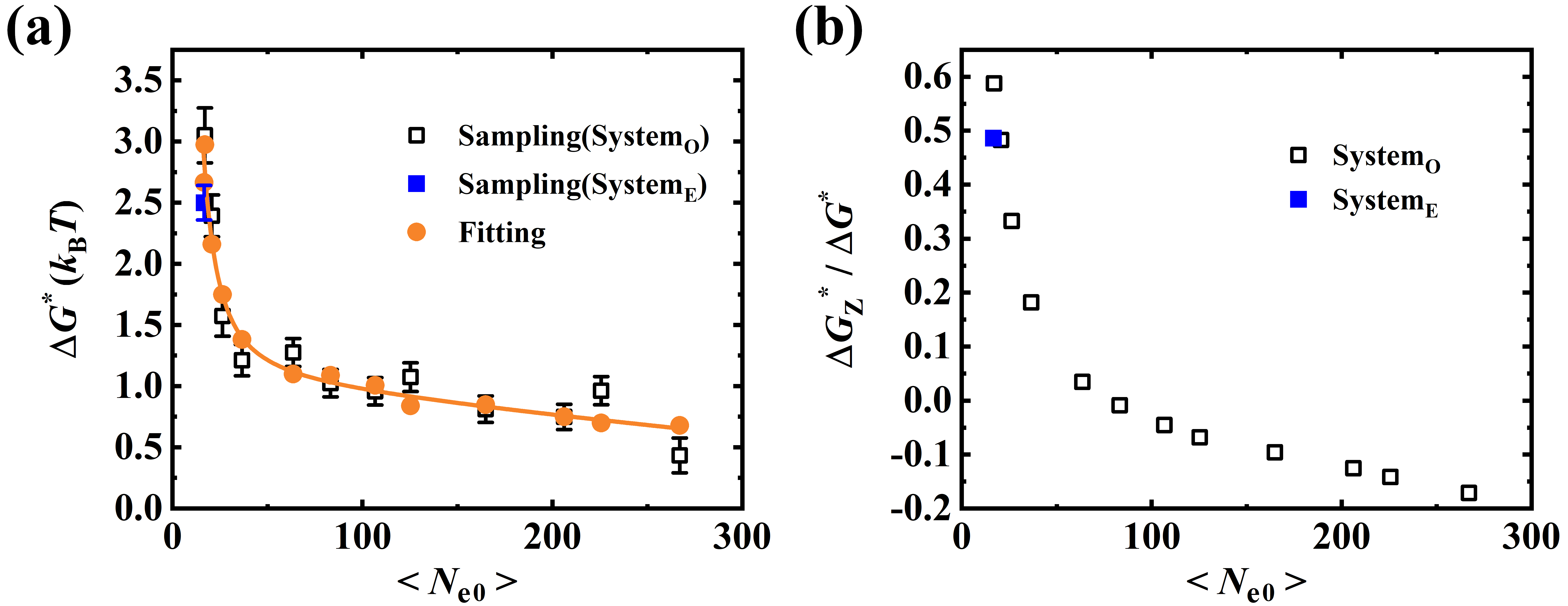}
  \caption{(a) Comparison of the nucleation free energy barriers calculated by sampling (open black squares) and fitted with Eq. (16) (orange circles). (b) The ratio of the entanglement induced nucleation free energy barrier to the total one, \(\Delta G_\mathrm{z}^*/\Delta G^*\) as a function of the entanglement length.}
  \label{f8}
\end{figure}

The ratio of the entanglement induced nucleation free energy barrier to the total one, \(\Delta G_\mathrm{z}^*/\Delta G^*\) can quantify the effect of the entanglement on the nucleation. As shown in Figure 8b, \(\Delta G_\mathrm{z}^*/\Delta G^* \approx 0.5\) for the equilibrium system, revealing that the entanglement makes a considerable impact on the nucleation. \(\Delta G_\mathrm{z}^*/\Delta G^*\) increases rapidly when \(\left \langle  N_{\mathrm{e0}}\right \rangle\) is smaller than the equilibrium value (ca. 17). This implies that the increased nucleation barrier is mainly caused by the denser entanglement. Indeed, our simulations show that the nucleation hardly takes place for systems with \(\left \langle  N_{\mathrm{e0}}\right \rangle = 6 \sim 15\). For systems with \(\left \langle  N_{\mathrm{e0}}\right \rangle\) being larger than the equilibrium value, \(\Delta G_\mathrm{z}^*/\Delta G^*\) decreases gradually with the increasing \(\left \langle  N_{\mathrm{e0}}\right \rangle\) due to the significantly increased entanglement entropy \(\Delta S_\mathrm{z}\). For systems with low entanglement densities, the formation of crystal stems increases constraints and drives the entanglement density close to the equilibrium value, which contributes a negative \(\Delta G_\mathrm{z}^*\) and consequently reduces the nucleation barrier predicted by the CNT.

The introduction of the entanglement free energy \(G_\mathrm{z}\) not only establishes a quantitative model to explain the entanglement effect in polymer crystallization experiments, but also provides a solution for the discrepancies in polymer crystallization. The difficulty in crystallization of highly entangled polymers like PC \cite{ref18,ref19}, the enhanced nucleation in cyclic polymers \cite{ref54,ref55} and disentangled systems with reduced entanglement density \cite{ref56,ref57,ref58} can be quantitatively correlated with \(G_\mathrm{z}\) via \(N_\mathrm{e}\). The discrepancies in the memory effect of crystallization \cite{ref22,ref23,ref59,ref60,ref61,ref62} and flow-induced crystallization of polymer \cite{ref63,ref64,ref65,ref66,ref67,ref68,ref69,ref70} might also be solved with \(G_\mathrm{z}\), where no theoretical consensus has been reached yet. In both cases, crystallization is accompanied by a higher initial \(G_\mathrm{z}\) and a greater increase in \(S_\mathrm{z}\), which makes the nucleation barrier lower than that of the equilibrium melt. In addition, the different interceptions of the polymer crystallization and melting lines measured in experiment \cite{ref71} may be partly caused by the different evolutionary pathways of the entanglement topology network during crystallization and melting, which makes the entanglement free energy change asymmetrical. Experimental and simulation efforts are encouraged to verify the quantitative correlation between entanglement and observations in polymer crystallization.

The entanglement free energy theory gives the theoretical expressions for entanglement in crystallization, which supplements the effect of the entanglement network in nucleation and diverts our attention from the chain conformation effect to the free energy of the entanglement network. Based on the entanglement free energy theory, the entanglement network has an intrinsic free energy \(G_\mathrm{z}\) independent of polymer conformation and structural configuration. The nucleation process brings additional topological entanglement points for the original entanglement network, which modifies \(G_\mathrm{z}\) that lives in the entanglement network. Therefore, during the nucleation process, not only the free energy of monomers inside a crystal nucleus is changed, but also the free energy of the neighboring entanglement network around the crystal nucleus is affected. Considering the chain connectivity, nucleation is no longer a local behavior among monomers, but a collaborative behavior between the nucleus and the entanglement network.

The concept of entanglement free energy proposed in this work builds a bridge between the entanglement network and the nucleation behavior of non-equilibrium entangled polymer systems. There are still several aspects to be improved in the future: (1) Strictly, \(U_\mathrm{z} \sim 1/N_\mathrm{e}^2\) is still a phenomenological theory. An analytical solution from the strict initio derivation at the molecular scale will be a serious challenge. (2) An improved expression for the entanglement entropy. The method to measure the two-body interactions between entanglement points remains to be perfected. One possible way is to calculate the entropy by counting the probability of microscopic entanglement states directly. (3) Verification of the entanglement free energy theory by experiments. Although the predictions by our theoretical approach can fit the simulation results well, a direct experimental evidence for the verification of our new theoretical approach is still necessary.

\section{4. Conclusions}
In conclusion, we have performed a series of MD simulations of polymer melts with different entanglement densities, and obtained the nucleation free energy barrier by using the modified MFPT method. A monotonic increase of the nucleation free energy barrier with the increasing entanglement density is observed, which is opposite to the prediction of the CNT. Then, we propose a new theoretical approach to take into account the unique chain connectivity of polymers. Specifically, we introduce the entanglement free energy to reflect the role of entanglement in polymer nucleation. It is suggested that polymer nucleation not only involves free energies of monomers inside and on the surface of a nucleus as considered in the CNT, but also affects the entanglement network around the nucleus. The calculated values of the nucleation free energy barrier based on our theoretical approach match well with those obtained from the simulations. Our study here provides new insights on how the intrinsic entanglement in polymer melts affects the nucleation behavior. 

\section{Associated content}
\subsection{Supporting Information}
The system preparation, crystallite identification criterion, nucleation rate, the introduction of MFPT method and data fitting process, the conformational entropy calculation, the atomic entropy calculation, the fold surface free energy calculation, and the derivation process of entanglement entropy expression.

\section{Acknowledgments}
We would like to thank Prof. Murugappan Muthukumar (UMass) for valuable discussions about how to compute surface free energy and Prof. Martin Kr\"{o}ger (Eidgen\"{o}ssische Technische Hochschule Z\"{u}rich) for his support with the Z1 algorithm. This work is supported by the National Key R\&D Program of China (2020YFA0405800), the National Natural Science Foundation of China (51890872, 51633009), and the Anhui Provincial Key R\&D Program (202004a05020075, 202104a05020008).

\bibliographystyle{unsrt}
\bibliography{article}

\begin{thebibliography}{10}

\bibitem{ref1}
J.~D. Hoffman and R.~L. Miller.
\newblock Miller, r. l. kinetic of crystallization from the melt and chain
  folding in polyethylene fractions revisited: theory and experiment.
\newblock {\em Polymer}, 38(13):3151--3212, 1997.

\bibitem{ref2}
Takashi Yamamoto.
\newblock Molecular dynamics simulation of polymer ordering. ii.
  crystallization from the melt.
\newblock {\em The Journal of Chemical Physics}, 115(18):8675--8680, 2001.

\bibitem{ref3}
S.~Shirvanyants, D.and~Panyukov, Q.~Liao, and M.~Rubinstein.
\newblock Long-range correlations in a polymer chain due to its connectivity.
\newblock {\em Macromolecules}, 41(4):1475--1485, 2008.

\bibitem{ref4}
T.~Yamamoto.
\newblock Molecular dynamics simulations of steady-state crystal growth and
  homogeneous nucleation in polyethylene-like polymer.
\newblock {\em J. Chem. Phys.}, 129(18):184903, 2008.

\bibitem{ref5}
T.~Yamamoto.
\newblock Molecular dynamics simulations of steady-state crystal growth and
  homogeneous nucleation in polyethylene-like polymer.
\newblock {\em J. Chem. Phys.}, 107(7):2653--2663, 1997.

\bibitem{ref6}
P.~G. de~Gennes.
\newblock Reptation of a polymer chain in the presence of fixed obstacles.
\newblock {\em J. Chem. Phys.}, 107(2):572--579, 1971.

\bibitem{ref7}
Richard~S. Graham, Alexei~E. Likhtman, Tom C.~B. McLeish, and Scott~T. Milner.
\newblock Microscopic theory of linear, entangled polymer chains under rapid
  deformation including chain stretch and convective constraint release.
\newblock {\em J. Rheol.}, 47(5):1171--1200, 2003.

\bibitem{ref8}
L.~Dai.
\newblock Developing the tube theory for polymer knots.
\newblock {\em Phys. Rev. Res.}, 2(2):022014, 2020.

\bibitem{ref9}
D.~Diddens, H.~Meyer, and A.~Johner.
\newblock Local chain segregation and entanglements in a confined polymer melt.
\newblock {\em Phys. Rev. Lett.}, 118(6):067802, 2017.

\bibitem{ref10}
P.~Welch and M.~Muthukumar.
\newblock Molecular mechanisms of polymer crystallization from solution.
\newblock {\em Phys. Rev. Lett.}, 87(21):218302, 2001.

\bibitem{ref11}
X.~Tang, W.~Chen, and L.~Li.
\newblock The tough journey of polymer crystallization: Battling with chain
  flexibility and connectivity.
\newblock {\em Macromolecules}, 52(10):3575--3591, 2019.

\bibitem{ref12}
E.~A. DiMarzio, C.~M. Guttman, and J.~D. Hoffman.
\newblock Is crystallization from the melt controlled by melt viscosity and
  entanglement effects?
\newblock {\em Faraday Discuss. Chem. Soc.}, 68(0):210--217, 1979.

\bibitem{ref13}
C.~Luo and J.-U. Sommer.
\newblock Disentanglement of linear polymer chains toward unentangled crystals.
\newblock {\em ACS Macro Letters}, 2(1):31--34, 2012.

\bibitem{ref14}
M.~Hikosaka, K.~Amano, S.~Rastogi, and A.~Keller.
\newblock Lamellar thickening growth of an extended chain single crystal of
  polyethylene. 1. pointers to a new crystallization mechanism of polymers.
\newblock {\em Macromolecules}, 30(7):2067--2074, 1997.

\bibitem{ref15}
D.~Lippits, S.~Rastogi, S.~Talebi, and C.~Bailly.
\newblock Formation of entanglements in initially disentangled polymer melts.
\newblock {\em Macromolecules}, 39(7):8882--8885, 2006.

\bibitem{ref16}
D.~R. Lippits, S.~Rastogi, G.~W.~H. Höhne, B.~Mezari, and P.~C. M.~M. Magusin.
\newblock Heterogeneous distribution of entanglements in the polymer melt and
  its influence on crystallization.
\newblock {\em Macromolecules}, 40(4):1004--1010, 2007.

\bibitem{ref17}
B.~Wunderlich.
\newblock NewYork, vol.ii edition, 1976.

\bibitem{ref18}
A.~Alizadeh, S.~Sohn, J.~Quinn, H.~Marand, L.~C. Shank, and H.~D. Iler.
\newblock Influence of structural and topological constraints on the
  crystallization and melting behavior of polymers: 3. bispheno a
  polycarbonate.
\newblock {\em Macromolecules}, 34(12):4066--4078, 2001.

\bibitem{ref19}
P.~R. Sundararajan.
\newblock Conformational features of bisphenol-a polycarbonate.
\newblock {\em Can. J. Chem.}, 63:103--110, 1985.

\bibitem{ref20}
K.~Liu, E.~L. de~Boer, Y.~Yao, D.~Romano, S.~Ronca, and S.~Rastogi.
\newblock Heterogeneous distribution of entanglements in a nonequilibrium
  polymer melt of uhmwpe: Influence on crystallization without and with
  graphene oxide.
\newblock {\em Macromolecules}, 49(19):7497--7509, 2016.

\bibitem{ref21}
S.~Yamazaki, M.~Hikosaka, A.Toda, I.~Wataoka., and F.~Gu.
\newblock Role of entanglement in nucleation and ‘melt relaxation’ of
  polyethylene.
\newblock {\em Polymer}, 43:6585--6593, 2002.

\bibitem{ref22}
C.~Luo and J.~U. Sommer.
\newblock Frozen topology: entanglements control nucleation and crystallization
  in polymers.
\newblock {\em Phys. Rev. Lett.}, 112(19):195702, 2014.

\bibitem{ref23}
C.~Luo and J.-U. Sommer.
\newblock Role of thermal history and entanglement related thickness selection
  in polymer crystallization.
\newblock {\em ACS Macro Letters}, 5(1):30--34, 2015.

\bibitem{ref24}
H.~G. Zachmann.
\newblock Statistische thermodynamik des kristallisierensund schmelzens von
  hochpolymeren stoffen.
\newblock {\em Kolloid Z. Z. Polymer}, 231(1−2):504--534, 1969.

\bibitem{ref25}
J.~D. Hoffman.
\newblock Role of reptation in the rate of crystallization of polyethylene
  fractions from the melt.
\newblock {\em Polymer}, 23(5):656--670, 1982.

\bibitem{ref26}
S.~Plimpton, P.~Crozier, and A.~Thompson.
\newblock Lammps-large-scale atomic/molecular massively parallel simulator.
\newblock {\em Sandia Natl. Lab}, 18, 2007.

\bibitem{ref27}
C.~Luo and J.-U. Sommer.
\newblock Coding coarse grained polymer model for lammps and its application to
  polymer crystallization.
\newblock {\em Comput. Phys. Commun.}, 180(8):1382--1391, 2009.

\bibitem{ref28}
M.~Kröger.
\newblock Shortest multiple disconnected path for the analysis of entanglements
  in two- and three-dimensional polymeric systems.
\newblock {\em Comput. Phys. Commun.}, 168(3):209--232, 2005.

\bibitem{ref29}
R.~S. Hoy, K.~Foteinopoulou, and M.~Kröger.
\newblock Topological analysis of polymeric melts: Chain-length effects and
  fast-converging estimators for entanglement length.
\newblock {\em Physical Review E}, 80(3):031803, 2009.

\bibitem{ref30}
N.~C. Karayiannis and M.~Kröger.
\newblock Combined molecular algorithms for the generation, equilibration and
  topological analysis of entangled polymers: Methodology and performance.
\newblock {\em International Journal of Molecular Sciences}, 10(11):5054--5089,
  2009.

\bibitem{ref31}
R.~S. Hoy and M.~Kröger.
\newblock Unified analytic expressions for the entanglement length, tube
  diameter, and plateau modulus of polymer melts.
\newblock {\em Phys. Rev. Lett.}, 124(14):147801, 2020.

\bibitem{ref32}
Ref [13] gives the equilibrium entanglement length of the original pva melt is
  around 25 monomers, while our measured value is 16 ~ 19. the value we get
  with Z1 is smaller than the value obtained by the previous version of Z1,
  here we attribute it to the systematic error.

\bibitem{ref33}
C.~L. Kelchner, S.~J. Plimpton, and J.~C. Hamilton.
\newblock Dislocation nucleation and defect structure during surface
  indentation.
\newblock {\em Phys. Rev. B}, 58:11085, 1998.

\bibitem{ref34}
C.~Luo and J.-U. Sommer.
\newblock Growth pathway and precursor states in single lamellar
  crystallization: Md simulations.
\newblock {\em Macromolecules}, 44(6):1523--1529, 2011.

\bibitem{ref35}
D.~A. Nicholson and G.~C. Rutledge.
\newblock Analysis of nucleation using mean first-passage time data from
  molecular dynamics simulation.
\newblock {\em J. Chem. Phys.}, 144(13):134105, 2016.

\bibitem{ref36}
D.~A. Nicholson and G.~C. Rutledge.
\newblock Molecular simulation of flow-enhanced nucleation in n-eicosane melts
  under steady shear and uniaxial extension.
\newblock {\em J. Chem. Phys.}, 145(24):244903, 2016.

\bibitem{ref37}
M.~Muthukumar.
\newblock Nucleation in polymer crystallization.
\newblock {\em Adv. Chem. Phys.}, 128:1--64, 2004.

\bibitem{ref38}
J.~I. Lauritzen and J.~D. Hoffman.
\newblock Theory of formation of polymer crystals with folded chains in dilute
  solution.
\newblock {\em J. Res. Natl. Bur. Stand., Sect. A}, 64A(1):73--102, 1960.

\bibitem{ref39}
J.~Krajenta, M.~Polińska, G.~Lapienis, and A.~Pawlak.
\newblock The crystallization of poly(ethylene oxide) with limited density of
  macromolecular entanglements.
\newblock {\em Polymer}, 197:122500, 2020.

\bibitem{ref40}
N.~Tian, D.~Liu, H.~Wei, Y.~Liu, and J.~Kong.
\newblock Crystallization of polycaprolactone with reduced entanglement.
\newblock {\em Eur. Polym. J.}, 102:38--44, 2018.

\bibitem{ref41}
D.~Turnbull and J.~C. Fisher.
\newblock Rate of nucleation in condensed systems.
\newblock {\em J. Chem. Phys.}, 17(1):71--73, 1949.

\bibitem{ref42}
P.~J. Flory.
\newblock Thermodynamics of crystallization in high polymers. i.
  crystallization induced by stretching.
\newblock {\em J. Chem. Phys.}, 15(6):397--408, 1947.

\bibitem{ref43}
P.~M. Piaggi, O.~Valsson, and M.~Parrinello.
\newblock Enhancing entropy and enthalpy fluctuations to drive crystallization
  in atomistic simulations.
\newblock {\em Phys. Rev. Lett.}, 119(1):015701, 2017.

\bibitem{ref44}
M.~H. Nafar~Sefiddashti, B.~J. Edwards, and B~Khomami.
\newblock A thermodynamically inspired method for quantifying phase transitions
  in polymeric liquids with application to flow-induced crystallization of a
  polyethylene melt.
\newblock {\em Macromolecules}, 53(23):10487--10502, 2020.

\bibitem{ref45}
R.~E. Nettleton and M.~S. Green.
\newblock Expression in terms of molecular distribution functions for the
  entropy density in an infinite system.
\newblock {\em J. Chem. Phys.}, 29:1365, 1958.

\bibitem{ref46}
M.~Muthukumar.
\newblock Molecular modelling of nucleation in polymers.
\newblock {\em Phil. Trans. R. Soc. A.}, 361(1804):539--556, 2003.

\bibitem{ref47}
M.~Andreev and G.~C. Rutledge.
\newblock A slip-link model for rheology of entangled polymer melts with
  crystallization.
\newblock {\em J. Rheol.}, 64(1):213--222, 2020.

\bibitem{ref48}
S.~Shanbhag, R.~G. Larson, J.~Takimoto, and M.~Doi.
\newblock Deviations from dynamic dilution in the terminal relaxation of star
  polymers.
\newblock {\em Phys. Rev. Lett.}, 87(19):195502, 2001.

\bibitem{ref49}
A.~Dambal, A.~Kushwaha, and E.~S.~G. Shaqfeh.
\newblock Slip-link simulations of entangled, finitely extensible, wormlike
  chains in shear flow.
\newblock {\em Macromolecules}, 42(18):7168--7183, 2009.

\bibitem{ref50}
S.~Shanbhag and R.~G. Larson.
\newblock A slip-link model of branch-point motion in entangled polymers.
\newblock {\em Macromolecules}, 37(21):8160--8166, 2004.

\bibitem{ref51}
K.~Iwata.
\newblock Role of entanglement in crystalline polymers 1. basic theory.
\newblock {\em Polymer}, 43:6609–6626, 2002.

\bibitem{ref52}
K.~Iwata and S.~F. Edwards.
\newblock New model of polymer entanglement: Localized gauss integral model.
  plateau modulus \(g_n\), topological second virial coefficient
  \(a_2^{\theta}\) and physical foundation of the tube model.
\newblock {\em J. Chem. Phys.}, 90(8):4567--4581, 1989.

\bibitem{ref53}
P.~M. Piaggi and M.~Parrinello.
\newblock Predicting polymorphism in molecular crystals using orientational
  entropy.
\newblock {\em Proc. Natl. Acad. Sci. U. S. A.}, 115(41):10251--10256, 2018.

\bibitem{ref54}
K.~Iyer and M.~Muthukumar.
\newblock Langevin dynamics simulation of crystallization of ring polymers.
\newblock {\em J. Chem. Phys.}, 148(24):244904, 2018.

\bibitem{ref55}
H.~Xiao, C.~Luo, D.~Yan, and J.-U. Sommer.
\newblock Molecular dynamics simulation of crystallization cyclic polymer melts
  as compared to their linear counterparts.
\newblock {\em Macromolecules}, 50(24):9796--9806, 2017.

\bibitem{ref56}
M.~Psarski, E.~Piorkowska, and A.~Galeski.
\newblock Crystallization of polyethylene from melt with lowered chain
  entanglements.
\newblock {\em Macromolecules}, 33(3):916--932, 2000.

\bibitem{ref57}
Z.~Zhai, C.~Fusco, J.~Morthomas, M.~Perez, and O.~Lame.
\newblock Disentangling and lamellar thickening of linear polymers during
  crystallization: Simulation of bimodal and unimodal molecular weight
  distribution systems.
\newblock {\em ACS Nano}, 13(10):11310--11319, 2019.

\bibitem{ref58}
T.~Yamamoto.
\newblock Crystallization of helical oligomers with chirality selection. i. a
  molecular dynamics simulation for bare helix.
\newblock {\em J. Chem. Phys.}, 125(6):64902, 2006.

\bibitem{ref59}
L.~Sangroniz, D.~Cavallo, and A.~J. Müller.
\newblock Self-nucleation effects on polymer crystallization.
\newblock {\em Macromolecules}, 53(12):4581--4604, 2020.

\bibitem{ref60}
F.~Su, X.~Li, W.~Zhou, S.~Zhu, Y.~Ji, Z.~Wang, Z.~Qi, and L.~Li.
\newblock Direct formation of isotactic poly(1-butene) form i crystal from
  memorized ordered melt.
\newblock {\em Macromolecules}, 46(18):7399--7405, 2013.

\bibitem{ref61}
M.~Muthukumar.
\newblock Communication: Theory of melt-memory in polymer crystallization.
\newblock {\em J. Chem. Phys.}, 145(3):031105, 2016.

\bibitem{ref62}
L.~Sangroniz, D.~Cavallo, A.~Santamaria, A.~J. Müller, and R.~G. Alamo.
\newblock Thermorheologically complex self-seeded melts of propylene–ethylene
  copolymers.
\newblock {\em Macromolecules}, 50(2):642--651, 2017.

\bibitem{ref63}
R.~S. Graham and P.~D. Olmsted.
\newblock Coarse-grained simulations of flow-induced nucleation in
  semicrystalline polymers.
\newblock {\em Phys. Rev. Lett.}, 103(11):115702, 2009.

\bibitem{ref64}
T.~Yamamoto.
\newblock Molecular dynamics simulation of stretch-induced crystallization in
  polyethylene: Emergence of fiber structure and molecular network.
\newblock {\em Macromolecules}, 52(4):1695--1706, 2019.

\bibitem{ref65}
D.~J. Read, C.~McIlroy, C.~Das, O.~G. Harlen, and R.~S. Graham.
\newblock Polystrand model of flow-induced nucleation in polymers.
\newblock {\em Phys. Rev. Lett.}, 124(14):147802, 2020.

\bibitem{ref66}
P.-A. Albouy and P.~Sotta.
\newblock Draw ratio at the onset of strain-induced crystallization in
  cross-linked natural rubber.
\newblock {\em Macromolecules}, 53(3):992--1000, 2020.

\bibitem{ref67}
K.~Cui, D.~Liu, Y.~Ji, N.~Huang, Z.~Ma, Z.~Wang, F.~Lv, H.~Yang, and L.~Li.
\newblock Nonequilibrium nature of flow-induced nucleation in isotactic
  polypropylene.
\newblock {\em Macromolecules}, 48(3):694--699, 2015.

\bibitem{ref68}
I.-C. Yeh, J.~W. Andzelm, and G.~C. Rutledge.
\newblock Mechanical and structural characterization of semicrystalline
  polyethylene under tensile deformation by molecular dynamics simulations.
\newblock {\em Macromolecules}, 48(12):4228--4239, 2015.

\bibitem{ref69}
S.~Jabbari-Farouji, J.~Rottler, O.~Lame, A.~Makke, M.~Perez, and J.-L. Barrat.
\newblock Plastic deformation mechanisms of semicrystalline and amorphous
  polymers.
\newblock {\em ACS Macro Letters}, 4(2):147--150, 2015.

\bibitem{ref70}
W.~Zhang and R.~G. Larson.
\newblock Effect of flow-induced nematic order on polyethylene crystal
  nucleation.
\newblock {\em Macromolecules}, 53(18):7650--7657, 2020.

\bibitem{ref71}
G.~Strobl.
\newblock Colloquium: Laws controlling crystallization and melting in bulk
  polymers.
\newblock {\em Rev. Mod. Phys.}, 81(3):1287--1300, 2009.

\end{thebibliography}

\end{document}